\documentclass[]{article}  

\usepackage[]{graphicx}
\usepackage{amsmath}
\usepackage{amssymb}
\usepackage{multirow}
\usepackage{subfigure}
\usepackage{epsfig}
\usepackage{wrapfig}
\usepackage{color}
\usepackage{cite}

\def\##1{{\bf #1}}
\def\=#1{\underline{\underline #1}}
\def\4#1{\underline{\underline{\underline{\underline #1}}}}

\def\.{\mbox{ \tiny{$^\bullet$} }}

\def\le{\left(}
\def\ri{\right)}
\def\les{\left[}
\def\ris{\right]}
\def\lec{\left\{}
\def\ric{\right\}}

\def\c#1{\cite{#1}}

\def\r#1{(\ref{#1})}
\def\rr#1{\ref{#1}}

\def\epso{\varepsilon_{\scriptscriptstyle 0}}
\def\lambdao{\lambda_{\scriptscriptstyle 0}}
\def\muo{\mu_{\scriptscriptstyle 0}}
\def\ko{k_{\scriptscriptstyle 0}}

\def\eps{\varepsilon}
\def\epsa{\varepsilon_a}
\def\epsb{\varepsilon_b}
\def\epsc{\varepsilon_c}
\def\epsd{\varepsilon_d}

\def\pinc{{\mathbf p}_+}
\def\pref{{\mathbf p}_-}

\def\aL{a_L}
\def\aR{a_R}
\def\rL{r_L}
\def\rR{r_R}

\def\inc{_{\rm inc}}
\def\refl{_{\rm ref}}

\def\ux{\hat{\mathbf{u}}_x}
\def\uy{\hat{\mathbf{u}}_y}
\def\uz{\hat{\mathbf{u}}_z}

\begin{document}
\pagestyle{plain}

  \begin{center}
{\bf Exhibition of circular Bragg phenomenon  by hyperbolic, dielectric, structurally chiral materials}\\

{\it Akhlesh Lakhtakia}\\

{
National Taipei University of Technology, Department of Electro-Optical Engineering,  Taipei 106, Taiwan\\
and\\
Pennsylvania State University, Department of Engineering Science and
Mechanics, 
University Park, Pennsylvania  16802--6812, USA\\
}
\end{center}

\begin{abstract}
The relative permittivity dyadic of a dielectric structurally chiral material (SCM) varies helicoidally along a fixed direction; in consequence, the SCM exhibits  the circular Bragg phenomenon, which is the circular-polarization-selective reflection
of light. The introduction of hyperbolicity in an SCM---by making either one or two but not all three eigenvalues of the
relative permittivity dyadic acquire negative real parts---does not eliminate the   circular Bragg phenomenon,
but significantly alters the regime for its exhibition.
Physical vapor deposition techniques appear to be suitable  to fabricate hyperbolic SCMs.

\noindent{\bf Keywords:}  circular Bragg phenomenon, hyperbolic dispersion, indefinite permittivity, structural chirality
 
\end{abstract}

\section{Introduction}

Chiral liquid crystals \c{Chan,deG} and chiral sculptured thin films \c{YK59,LMbook} are dielectric examples of structurally chiral materials (SCMs)---which are anisotropic 
and helicoidally nonhomogeneous along a fixed axis. If that fixed axis is parallel to the $z$ axis of a Cartesian coordinate
system with unit vectors $\ux$, $\uy$, and $\uz$, the frequency-domain constitutive relations of dielectric SCMs may be
set down as
\begin{equation}
\label{conrel}
\left. \begin{array}{l}
\#D(\#r)=\epso\,\=\eps_r(z)\.\#E(\#r)\\[5pt]
\qquad= \epso \,\=S_z(z)\.
 \=S_y(\chi)\. \left(
\epsa  \, \uz\uz+ \epsb \, \ux\ux
\right.  \\ \qquad\qquad \left.
+\, \epsc \, \uy\uy\right)
\.\=S_y^T(\chi)\.\=S_z^T(z)\.\#E(\#r) \\[5pt]
\#B(\#r)= \muo\,  \#H(\#r)\,,
\end{array}\right\}\,,
\end{equation}
where the superscript T denotes the transpose;  $\muo$ and $\epso$ are  the permeability and permittivity of
free space; the rotational nonhomogeneity  is expressed through the
dyadic
\begin{equation}
\=S_z(z)=\uz\uz +\le\ux\ux+\uy\uy\ri\cos\le\frac {h\pi z}{\Omega}\ri
+\le\uy\ux-\ux\uy\ri\sin\le\frac {h\pi z}{\Omega}\ri\,,
\label{Sz-def}
\end{equation}
with $2\Omega$ as the helical pitch and either $h=+1$ for structural right-handedness
or $h=-1$ for structural left-handedness; the dyadic
\begin{equation}
\=S_y(\chi)=\le \ux\ux + \uz\uz \ri \cos{\chi}
+\le \uz\ux -
\ux\uz \ri \sin{\chi}+\uy\uy 
\label{Sy-def}
\end{equation}
containing $\chi\in[0^\circ,90^\circ]$ as the angle of rise of the helical morphology; and
$\epsa$, $\epsb$, and $\epsc$ are the three $z$-independent eigenvalues of the relative permittivity
dyadic $\=\eps_r(z)$. 
Typically, dissipation is small enough to be ignored and $\eps_{a,b,c}>0$; hence, $\=\eps_r(z)$ is positive
definite \c{Lut}. Figure~\rr{Fig1} shows a cross-sectional image of a chiral sculptured thin film.

The optical signature of an SCM is a circular-polarization-sensitive stopband. The center wavelength and
the width of this stopband depend on the direction of the wavevector of an incident circularly polarized
plane wave. Most
significantly, the stopband is exhibited  when the incident plane wave's handedness is the same as the
structural handedness of the SCM, but not otherwise. The stopband is best seen when the thickness of the SCM exceeds
a certain number of helical pitches \c{Fergason,StJohn,WHL00,HWTLM}. When dissipation is small enough to be ignored,
 $\=\eps_r(z)$ is positive definite, and the variations of $\eps_{a,b,c}$ with respect to the free-space
 wavelength $\lambdao$ are also small enough to be ignored,
 the circular-polarization-sensitive stopband can be delineated as \c{LMbook}
\begin{equation}
\frac{\lambdao}{2\Omega } \in 
\left\{
\begin{array}{ll}
\les \sqrt{\epsc},\sqrt{\epsd}\ris \cos^{1/2}\theta\,, &\qquad \epsc < \epsd\,,
\\[5pt]
\les \sqrt{\epsd},\sqrt{\epsc}\ris \cos^{1/2}\theta\,, &\qquad \epsc > \epsd\,,
\end{array}
\right.
\label{limits}
\end{equation}
where $\theta$ is the angle of incidence with respect to the $z$ axis and
\begin{equation}
\epsd= \frac{\epsa\epsb}{\epsa\,\cos^2\chi+\epsb\,\sin^2\chi}\,.
\end{equation}
Provided that ${\rm Re}\les\eps_\sigma\ris\gg\vert{\rm Im}\les\eps_\sigma\ris\vert$ for all
$\sigma\in\lec{a,b,c}\ric$, the estimates \r{limits} can be used with $\epsc$ replaced
by $\vert\epsc\vert$ and $\epsd$ by $\vert\epsd\vert$. The exhibition of
the circular-polarization-sensitive stopband is called the circular Bragg phenomenon.

During the last ten years, attention has been paid to dielectric-magnetic
materials with indefinite  permeability and
permittivity dyadics \c{Smith,DIL}. Although the practical realization of such materials 
remains a matter of conjecture, there is no doubt on the existence in nature of dielectric materials 
 the real parts of whose permittivity dyadics are indefinite \c{Lut}. Graphite \c{Sun}, triglycine sulfate 
\c{Gerbaux,Alekseyev}, sapphire \c{Schubert}, and bismuth \c{Alekseyev}
are examples.
Metal nanowire arrays \c{Kanungo} and periodic metal/dielectric multilayers \c{Kidwai}
provide examples of manufactured anisotropic dielectric materials
whose effective permittivity dyadics have indefinite real parts \c{Cortes}. Periodic graphene/dielectric multilayers
have also been profferred as candidates \c{Othman}. Although dissipation due to conduction in metals
and graphene has been predicted to be offset-able by using dielectric materials with optical gain \c{Ni},
the   effective-medium approximations underlying such predictions must be handled with some care \c{MLoc2009}.

Experience with ambichiral materials \c{GY,HLWDM}
indicates that hyperbolic, dielectric SCMs ought to be practically realizable as
nanoengineered periodic multilayers. A variety of physical deposition techniques---such
as thermal evaporation, electron-beam evaporation, and sputtering \cite{MPLbook}---can be 
used to deposit alternating layers of a metal and a dielectric material on a suitably rotating planar substrate \c{LMbook}.
In these fabrication techniques, collimated vapor
fluxes of both materials must be directed very obliquely towards the substrate in order to engender biaxiality.
Furthermore,  the nominal thickness of each metal layer must be a small fraction of the nominal thickness
of each dielectric layer \c{MLoc2009}, and all layers must be electrically thin \c{King,Bohren}.
Either one or two of ${\rm Re}\left(\epsa\right)$, ${\rm Re}\left(\epsb\right)$,
and ${\rm Re}\left(\epsc\right)$ would be negative, with the remainder being positive. Then, with dissipation 
assumed to be sufficiently
small,  the  estimates \r{limits} would become dubious. Indeed, the question arises: will  a hyperbolic SCM exhibit the
circular Bragg phenomenon?  

In order to answer this question, a one-point boundary-value problem was formulated and solved. In this problem, the
half space $z<0$ is vacuous while the half space $z>0$ is occupied by the hyperbolic SCM, and a circularly polarized
plane wave is obliquely incident on the interface $z=0$ from its vacuous side. 
As it is known that the circular Bragg phenomenon develops as the thickness of an SCM increases \c{StJohn,LMbook},  
an SCM half space should conceptually deliver the best developed  circular Bragg phenomenon.
The underlying  boundary-value
 problem is  introduced briefly in Sec.~\ref{theory},
the detailed procedure to solve it being available elsewhere \c{LakhPLA2010}.
Numerical results are provided and discussed
in Sec.~\ref{nrd}. 
An $\exp(-i\omega t)$  dependence on time $t$ is implicit, with $\omega$
denoting the angular frequency and $i=\sqrt{-1}$. The free-space wavenumber is denoted by $\ko=\omega\sqrt{\epso\muo}=2\pi/\lambdao$.

\section{Boundary-value problem}\label{theory}
Let the half space $z<0$ be vacuous, while the half space $z>0$ be occupied by an SCM described
by Eqs.~\r{conrel}--\r{Sy-def}.
An arbitrarily polarized plane wave 
is obliquely incident 
on the interface $z=0$ from the vacuous half space.  Without significant loss of generality,
let the wave vector of this
 plane wave  lie wholly in the $xz$ plane and make
an angle $\theta  \in \les 0^\circ,90^\circ\ri$ with respect to the $+z$ axis. Accordingly, the electric field
phasor of the incident plane wave may be written as
\begin{equation}
\#E\inc=\le \aL\,\frac{i\uy-\pinc}{\sqrt{2}} -
\aR\,\frac{i\uy+\pinc}{\sqrt{2}} \ri\,\exp
\les{i\ko \le{x}\sin\theta +{z}\cos\theta \ri}\ris \,, \quad z\leq 0\,,
\label{Einc}
\end{equation}
where
$\aL$ and $\aR$ are the known
amplitudes of the  left- and right-circularly polarized   components, respectively, and
the vectors
\begin{equation}
 \#p_\pm=\mp \ux  \cos\theta  + 
\uz \sin\theta 
\end{equation}
are of unit magnitude.  
The reflected plane wave's electric field
phasor is given by
\begin{equation}
\#E\refl=\le -\rL\,\frac{i\uy-\pref}{\sqrt{2}} +
\rR\,\frac{i\uy+\pref}{\sqrt{2}} \ri\,\exp
\les{i\ko \le{x}\sin\theta -{z}\cos\theta \ri}\ris \,, \quad z\leq 0\,,
\label{Erefl}
\end{equation}
with unknown amplitudes $\rL$ and $\rR$. The procedure to determine
$\rL$ and $\rR$ in terms of $\aL$ and $\aR$ is described in detail elsewhere \c{LakhPLA2010}.

The reflection amplitudes are related to the incidence amplitudes
by the four reflection  coefficients
entering the 2$\times$2 matrix in the following relation:
\begin{equation}
\label{eq16}
\les \begin{array}{cccc} \rL\\ \rR  \end{array}\ris  =
\les \begin{array}{cccc} r_{LL} & r_{LR} \\ r_{RL} & r_{RR}\end{array}\ris \,
\les \begin{array}{cccc} \aL \\ \aR  \end{array}\ris
\, .
\end{equation}
The co-polarized reflectances of the SCM half space are denoted by $R_{LL}=\vert r_{LL}\vert^2$ and
$R_{RR} = \vert r_{RR}\vert^2$,
and the cross-polarized ones by $R_{LR}=\vert r_{LR}\vert^2$  and 
$R_{RL}=\vert r_{RL}\vert^2$. The principle of conservation of energy requires that
$R_R=R_{RR} +R_{LR} \leq 1$ and $R_L=R_{LL} +R_{RL} \leq 1$.

\section{Numerical results and discussion}\label{nrd}

Parametric calculations were made with the SCM assumed to be structurally right handed ($h=1$),
with all three of $\eps_{a,b,c}$ chosen to have very small and positive imaginary parts (that are indicative of weak dissipation).
The reflectances $R_{LL}$, $R_{RL}$, $R_{RR}$, and $R_{LR}$ were computed as functions of the angle of incidence
$\theta$
and either (i) the normalized wavelength  $\lambdao/2\Omega$ for fixed  angle of rise $\chi$ or
(ii) $\chi$ for fixed  $\lambdao/2\Omega$.

In order to set a baseline for discussion, Fig.~\ref{Fig2} displays all four reflectances
as functions of  the normalized wavelength  $\lambdao/2\Omega$ and the angle of incidence $\theta$,
when $\chi=60^\circ$ and the SCM is of the regular (i.e., non-hyperbolic) type: $\epsa=3.26(1+0.001i)$, $\epsb=4.46(1+0.001i)$, and
$\epsc=3.78(1+0.001i)$. A sigmoid ridge of high values of $R_{RR}$ is evident in the figure. 
The limits \r{limits} with $\eps_\sigma$ replaced by $\vert\eps_\sigma\vert$,
$\sigma\in\lec{a,b,c}\ric$, are satisfied by this ridge. For $\theta\stackrel{<}{\sim}70^\circ$,
$R_{LL}$ is negligible in that portion of the $(\lambdao/2\Omega)$-$\theta$ plane which is occupied by the
high-$R_{RR}$ ridge; additionally, both cross-polarized reflectances are  very small. The huge excess
of $R_{RR}$ over $R_{LL}$
accompanied by very small values of the two other reflectances is the
chief manifestation of the circular Bragg phenomenon. 

When the sign of ${\rm Re}\le\epsb\ri$ was altered from positive to negative, the SCM of Fig.~\ref{Fig2} became hyperbolic
and the sigmoid high-$R_{RR}$ ridge of that figure disappeared. However, a search with somewhat higher values
of $\lambdao/2\Omega$ soon revealed a portion of the $(\lambdao/2\Omega)$-$\theta$ plane in which (i) $R_{RR}$ exceeds
$R_{LL}$ by a large margin and (ii) the excess of $R_{R}$ over $R_{L}$ is even greater, with $R_{LL}$ very close
to zero. In Fig.~\ref{Fig3},
circular-polarization-selective reflection is clearly evident for $\lambdao/2\Omega\in[1.95,2.4]$ and $\theta\in[0^\circ,20^\circ]$
as well as in the upper left neighborhood of that rectangular zone, for the hyperbolic SCM.

Next, for the computation of the reflectances presented in Fig.~\ref{Fig4} as functions of  $\lambdao/2\Omega$
and $\theta$, the following
parameters were used: $\epsa=3.26(-1+0.001i)$, $\epsb=4.46(1+0.001i)$, 
$\epsc=3.78(-1+0.001i)$, and $\chi=15^\circ$. Thus two of  the three   eigenvalues of  $\=\eps_r(z)$ now have negative real parts.
Circular-polarization-selective reflection with $R_{LL}$ almost equal to zero  is evident in  Fig.~\ref{Fig4} for $\lambdao/2\Omega\in[2,2.2]$ and $\theta\in[0^\circ,70^\circ]$ as well as on the outskirts of this rectangular zone in the $(\lambdao/2\Omega)$-$\theta$ plane.

The exhibition of the circular Bragg phenomenon by a regular  SCM 
for fixed values of $\lambdao/2\Omega$ and $\theta$ is delineated by the expression \r{limits}. One just has to ensure
the appropriate choices of $\epsc$ and $\epsd$, the correct choice of the latter parameter being determined by the correct
choices of $\epsa$, $\epsb$, and $\chi$. If all three of the   eigenvalues of  $\=\eps_r(z)$ are fixed as well,
then an appropriate value of $\sin\chi$ and must be found. But no physical value of $\chi$ may emerge. Therefore,
the exhibition of circular-polarization-selective reflection by hyperbolic SCMs was investigated in the $\chi$-$\theta$ plane
for fixed values of  $\epsa$, $\epsb$, $\epsc$, and $\lambdao/2\Omega$.

Figure~\ref{Fig5} shows   all four reflectances computed as functions of $\chi\in[0^\circ,90^\circ]$ and $\theta\in[0^\circ,90^\circ)$, when $\epsa=3.26(1+0.001i)$, $\epsb=4.46(-1+0.001i)$, $\epsc=3.78(1+0.001i)$, and $\lambdao/2\Omega=1.918$. Thus, only one of the three eigenvalues of
$\=\eps_r(z)$ has a negative real part. At least four distinct zones of high values of $R_{RR}$ accompanied by almost-zero values
of $R_{LL}$ and very low values of both cross-polarized reflectances can be identified in this figure.

Similar data  computed for $\epsa=3.26(-1+0.001i)$, $\epsb=4.46(1+0.001i)$, and $\epsc=3.78(-1+0.001i)$ are displayed
in Fig.~\ref{Fig6}. Now, two of the three eigenvalues of
$\=\eps_r(z)$ have negative real parts. Circular-polarization-selective reflection with $R_{LL}\approx 0$ 
and very low values of $R_{LR}$ and $R_{RR}$ is evident for
$\chi\in[0^\circ,15^\circ]$ and $\theta\in[0^\circ,70^\circ]$.

To conclude, the concept of hyperbolic structurally chiral materials was introduced in this communication. The
hyperbolicity was found to significantly alter---but not eliminate---the exhibition of the circular Bragg phenomenon,
which has long been known to be the distinctive signature of non-hyperbolic SCMs such as cholesteric liquid
crystals \c{Fergason} and chiral sculptured thin films \c{WHL00}. Although practical realization of hyperbolic
SCMs has yet to occur, physical vapor deposition offers suitable   techniques to fabricate
 these materials.

\bigskip
\noindent {\bf Acknowledgment.}
The National Taipei University of Technology is thanked by the author
for an honorary international chair professorship.  The Charles Godfrey Binder Endowment at Penn State is thanked for
partial support of the author's research activities.

\newpage
\begin{figure}[ht]
\begin{center}
\resizebox*{4cm}{!}{\includegraphics{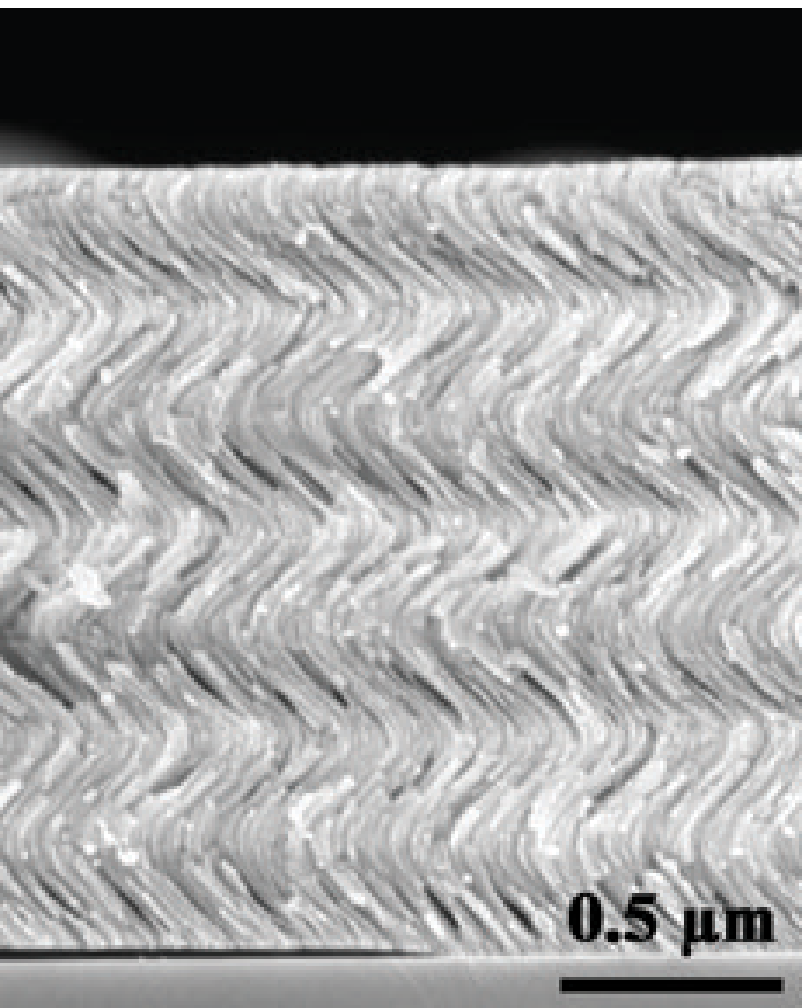}}%
\caption{Cross-sectional scanning electron micrograph of a chiral sculptured thin film made by thermal evaporation
of a chalcogenide glass.}%
\label{Fig1}
\end{center}
\end{figure}

\newpage 
\begin{figure}[ht]
\begin{center}
\resizebox*{8cm}{!}{\includegraphics{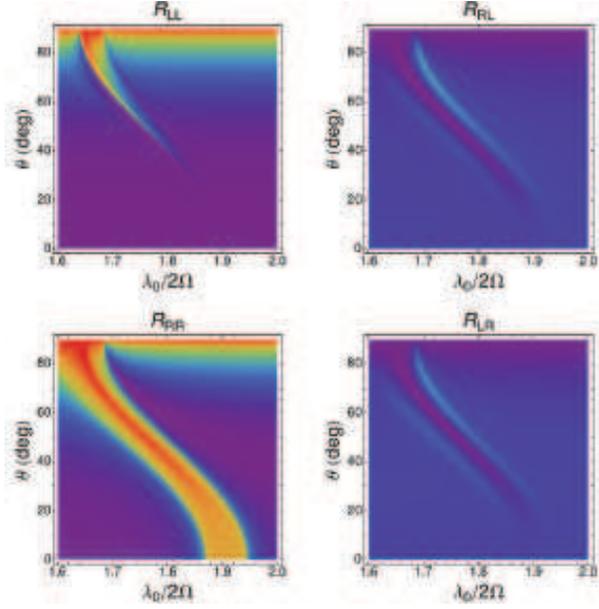}}%
\caption{Reflectances $R_{LL}$, $R_{RL}$, $R_{RR}$, and $R_{LR}$ as functions of $\lambdao/2\Omega$
and $\theta$, when $\epsa=3.26(1+0.001i)$, $\epsb=4.46(1+0.001i)$, $\epsc=3.78(1+0.001i)$, $\chi=60^\circ$,
and $h=1$. The color coding employs the spectrum of the rainbow with the deepest violet denoting $0$ and the 
deepest red
denoting $1.0$.
}%
\label{Fig2}
\end{center}
\end{figure}

\newpage 
\begin{figure}[ht]
\begin{center}
\resizebox*{8cm}{!}{\includegraphics{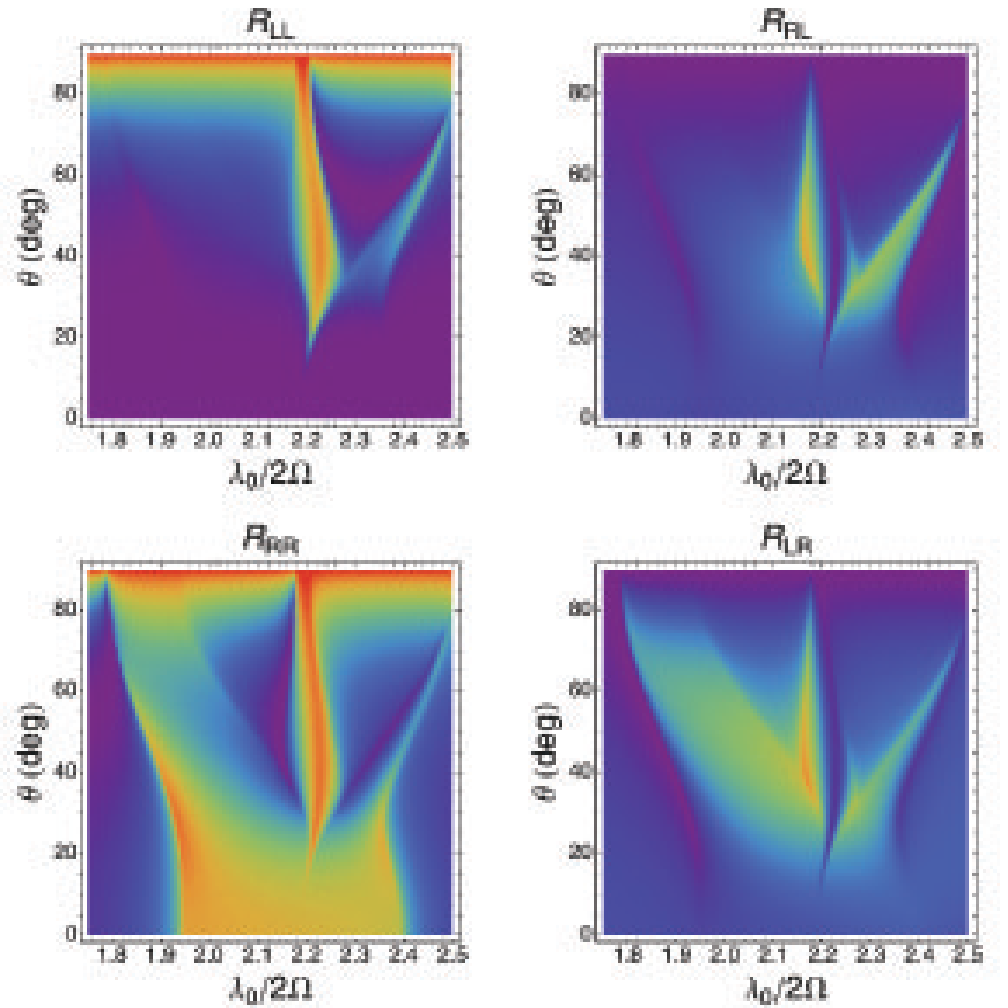}}%
\caption{Same as Fig.~\ref{Fig2}, except that $\epsb=4.46(-1+0.001i)$.
}%
\label{Fig3}
\end{center}
\end{figure}

\newpage 
\begin{figure}[ht]
\begin{center}
\resizebox*{8cm}{!}{\includegraphics{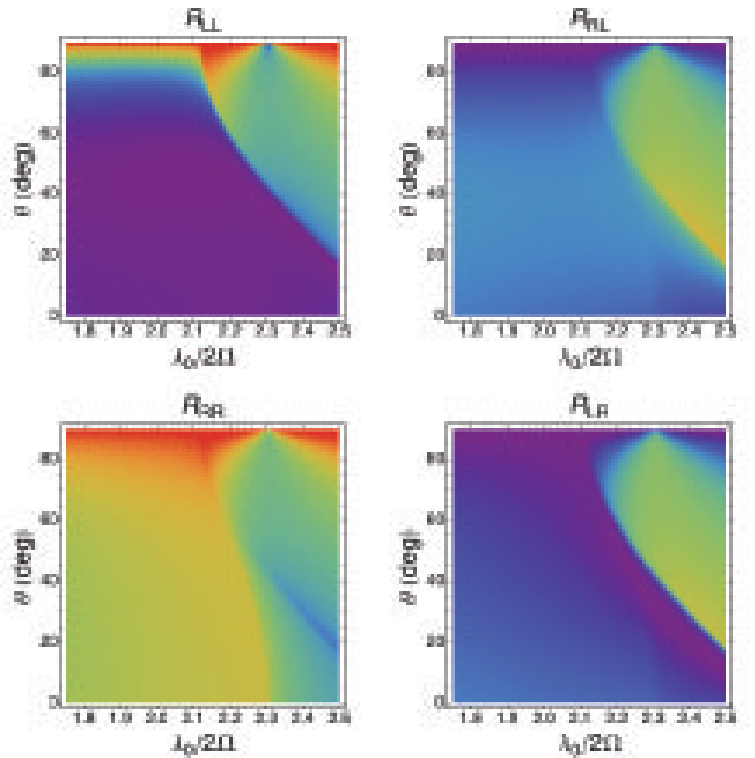}}%
\caption{Same as Fig.~\ref{Fig2}, except that $\epsa=3.26(-1+0.001i)$,  $\epsc=3.78(-1+0.001i)$, and $\chi=15^\circ$.
}%
\label{Fig4}
\end{center}
\end{figure}

\newpage 
\begin{figure}[ht]
\begin{center}
\resizebox*{8cm}{!}{\includegraphics{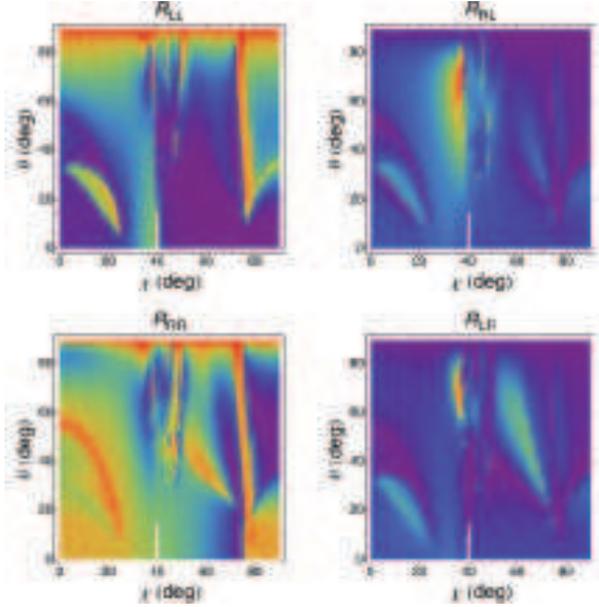}}%
\caption{Reflectances $R_{LL}$, $R_{RL}$, $R_{RR}$, and $R_{LR}$ as functions of $\chi$
and $\theta$, when $\epsa=3.26(1+0.001i)$, $\epsb=4.46(-1+0.001i)$, $\epsc=3.78(1+0.001i)$, $\lambdao/2\Omega=1.918$,
and $h=1$. The color coding employs the spectrum of the rainbow with the deepest violet denoting $0$ and the 
deepest red
denoting $1.0$. The thin white strips for $\theta<20^\circ$ and $\chi\approx40^\circ$ indicate a failure of the computational algorithm \c{LakhPLA2010}.
}%
\label{Fig5}
\end{center}
\end{figure}

\newpage 
\begin{figure}[ht]
\begin{center}
\resizebox*{8cm}{!}{\includegraphics{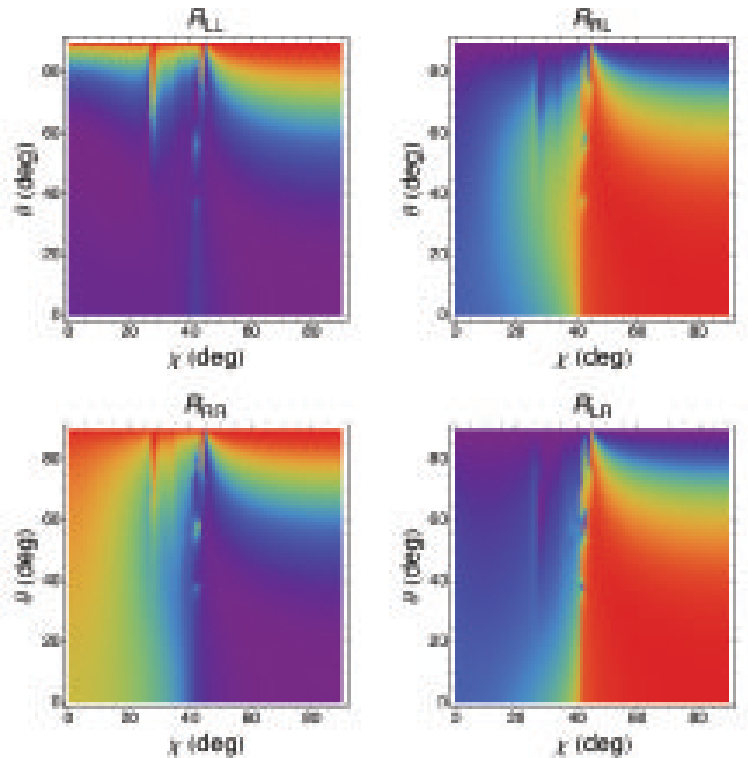}}%
\caption{Same as Fig.~\ref{Fig5}, except  $\epsa=3.26(-1+0.001i)$, $\epsb=4.46(1+0.001i)$, 
and $\epsc=3.78(-1+0.001i)$.
}%
\label{Fig6}
\end{center}
\end{figure}


\begin{thebibliography}{99}

 
\bibitem{Chan}
S. Chandrasekhar,  \textit{Liquid Crystals, 2nd ed.}, Cambridge University Press, Cambridge, United Kingdom (1992).
 
\bibitem{deG}
P. G. de Gennes and J. A. Prost,  \textit{The Physics of Liquid Crystals, 2nd ed.}, Clarendon Press,
Oxford, United Kingdom (1993).

\bibitem{YK59}
N. O. Young and J. Kowal, 
``Optically active fluorite films,"
 \textit{Nature} \textbf{183}(4654), 104--105 (1959),
{http://dx.doi.org/10.1038/183104a0}.
 
\bibitem{LMbook}
A. Lakhtakia and R. Messier,
{\it Sculptured Thin Films: Nano\-engineered Morphology and
Optics}, SPIE Press, Bellingham, WA, USA (2005), Chap.~9.

\bibitem{Lut}
H. L\"utkepohl,
{\it Handbook of Matrices}, Wiley, Chicester, United Kingdom (1996), Chap.~9.

\bibitem{Fergason}
J. L. Fergason, 
``Cholesteric structure---I Optical properties," 
\textit{Molec. Cryst.} {\bf 1}(2), 293--307 (1966),
{http://dx.doi.org/10.1080/15421406608083274}.

\bibitem{StJohn}
W. D. St. John,
W. J. Fritz, Z. J. Lu, and D.-K. Yang, 
``Bragg reflection from cholesteric liquid crystals,"
\textit{Phys. Rev. E} {\bf 51}(2), 1191--1198 (1995),
{http://dx.doi.org/10.1103/PhysRevE.51.1191}.

\bibitem{WHL00}
Q. Wu,  I. J. Hodgkinson, and A.  Lakhtakia,  
``Circular polarization filters made of chiral sculptured thin films: 
experimental and simulation results,"
\textit{Opt. Eng.} \textbf{39}(7), 1863--1868 (2000),
{http://dx.doi.org/10.1117/1.602570}.

\bibitem{HWTLM}
I. J. Hodgkinson,
Q. Wu, K. E. Thorn, A.  Lakhtakia, and M. W. McCall,  
``Spacerless circular-polarization spectral-hole filters using chiral 
sculptured thin films: theory and experiment,"
\textit{Opt. Commun.}  \textbf{184}(1-4), 57--66 (2000),
{http://dx.doi.org/10.1016/S0030-4018(00)00935-4}.

\bibitem{Smith}
D. R. Smith, P. Kolinko, and D. Schurig,
``Negative refraction
in indefinite media," \textit{J. Opt. Soc. Am. B} \textbf{21}, 1032--1043 (2004),
{http://dx.doi.org/10.1364/JOSAB.21.001032}.


\bibitem{DIL}
R. A. Depine, M. E. Inchaussandague, and A. Lakhtakia,
``Classification of dispersion equations for
homogeneous, dielectric-magnetic, uniaxial
materials," \textit{J. Opt. Soc. Am. A} {\bf 23}(4), 949--955 (2005), 
{http://dx.doi.org/10.1364/JOSAA.23.000949}.

\bibitem{Sun}
J. Sun, J. Zhou, B. Li, and F. Kang, ``Indefinite permittivity and
negative refraction in natural material: graphite," {\it Appl. Phys. Lett.} {\bf 98}(10),
101901 (2011), 
{http://dx.doi.org/10.1063/1.3562033}.

\bibitem{Gerbaux}
X. Gerbaux, M. Tazawa, and A. Hadni, ``Far IR transmission
measurements on triglycine sulphate (TGS), at 5K," \textit{Ferroelectrics}
{\bf 215}(1), 47--63 (1998),
{http://dx.doi.org/10.1080/00150199808229549}.

\bibitem{Alekseyev}
L. V. Alekseyev, V. A. Podolskiy, and E. E. Narimanov,
``Homogeneous hyperbolic systems for terahertz
and far-infrared frequencies," \textit{Adv. OptoElectron.} {\bf 2012}, 267564 (2000),
{http://dx.doi.org/10.1155/2012/267564}.

\bibitem{Schubert}
M. Schubert, T. E. Tiwald, and C. M. Herzinger, ``Infrared
dielectric anisotropy and phonon modes of sapphire," \textit{Phys.
Rev. B} \textbf{61}(12),  8187--8201 (2000),
{http://dx.doi.org/10.1103/PhysRevB.61.8187}.



\bibitem{Kanungo}
J. Kanungo and J. Schilling, ``Experimental determination of the principal
dielectric functions in silver nanowire metamaterials," \textit{Appl. Phys. Lett.} {\bf 97}(2),
021903 (2010),
{http://dx.doi.org/10.1063/1.3462311}.


\bibitem{Kidwai}
O. Kidwai, S. V. Zhukovsky, and J. E. Sipe, ``Effective-medium approach to planar multi\-layer
hyperbolic metamaterials: strengths and limitations," \textit{Phys. Rev. A} {\bf 85}(5), 053842
(2012), {http://dx.doi.org/10.1103/PhysRevA.85.053842}.

\bibitem{Cortes}
C. L. Cortes,
W. Newman, S. Molesky, and Z. Jacob, 
``Quantum nanophotonics
using hyperbolic metamaterials," \textit{J. Opt. (UK)} {\bf 14}(6), 063001 (2012), 
{http://dx.doi.org/10.1088/2040-8978/14/6/063001}.

\bibitem{Othman}
M. A. K. Othman, C. Guclu, and F. Capolino,
``Graphene-dielectric composite metamaterials: evolution
from elliptic to hyperbolic wavevector dispersion and
the transverse epsilon-near-zero condition," \textit{J. Nanophotonics}
{\bf 7}, 073089 (2013),
{http://dx.doi.org/10.1117/1.JNP.7.073089}.

\bibitem{Ni}
X. Ni,
S. Ishii, M.  D. Thoreson, V. M. Shalaev, S. Han,
S. Lee, and A. V. Kildishev,
``Loss-compensated and active hyperbolic
metamaterials," \textit{Opt. Express} {\bf 19}(25),
25242--25254 (2011),
{http://dx.doi.org/10.1364/OE.19.025242}.

\bibitem{MLoc2009}
T. G. Mackay and A. Lakhtakia,
``On the application of homogenization formalisms to active dielectric
composite materials,"
\textit{Opt. Commun.} {\bf 282}(13),
2470--2475 (2009),
{http://dx.doi.org/10.1016/j.optcom.2009.03.035}.

\bibitem{GY}
H. J. Gerritsen and R. T. Yamaguchi, 
``A microwave analog of optical rotation in cholesteric liquid crystals," 
\textit{Am. J. Phys.} {\bf 39}(8), 920--923 (1971),
{http://dx.doi.org/10.1119/1.1986325}.

 
\bibitem{HLWDM}
I. J. Hodgkinson,
A. Lakhtakia, Q. h. Wu, L. De Silva, and M. W. McCall, 
``Ambichiral, equichiral and finely chiral layered structures,"
\textit{Opt. Commun.} {\bf 239}(4-6), 353--358 (2004),
{http://dx.doi.org/10.1016/j.optcom.2004.06.005}.

\bibitem{MPLbook}
R.~J. Mart\'in-Palma and A. Lakhtakia,
{\it Nano\-technology: A Crash Course}, SPIE Press, Bellingham, WA, USA (2010), Chap.~4.

\bibitem{King}
R. W. P. King and S. S. Sandler,
``The electromagnetic field of a vertical electric dipole in the presence of a three-layered region,"
\textit{Radio Sci.} {\bf 29}(1), 97--113 (1994),
{http://dx.doi.org/10.1029/93RS02662}.

\bibitem{Bohren}
C. F. Bohren, X. Xiao, and A. Lakhtakia,
``The missing ingredient in effective-medium theories: standard deviations,"
\textit{J. Modern Opt.} {\bf 59}(15), 1312--1315 (2012),
{http://dx.doi.org/10.1080/09500340.2012.713521}.


\bibitem{LakhPLA2010}
A. Lakhtakia,
``Reflection of an obliquely incident plane wave by a half space
filled by a helicoidal bianisotropic medium,"
\textit{Phys. Lett. A}  \textbf{374}, 3887--3894 (2010),
{http://dx.doi.org/10.1016/j.physleta.2010.07.047}.

 

\end{thebibliography}
\end{document}